\documentstyle[12pt]{article}
\begin{document}
\title{THE MESON MODEL ON THE BASIS OF THE STRING SOLUTION OF THE
HEISENBERG EQUATION}
\author{V.D.Dzhunushaliev}
\date{}
\maketitle
\begin{center}
Theoretical physics department, the Kyrgyz State National
University, 720024, Bishkek, Kyrgyzstan
\end{center}
\begin{abstract}
\par
The axially symmetric non-local solution in the Heisenberg
equation is found.  It  is  regular  in  the whole space
and has the
finite energy on the unit of length according to this we may
consider the solution  as  a  string. Taking the non-local
spherically symmetric solution, which was found by
Finkelstein et. al.,
and our solution in account we suggest to  consider the
Heisenberg equation  as a quantum equation for non-local objects
(strings, flux tubes, membranes and so on). The received
solution is used for the obtaining the meson model as a
rotating string with the quark on its ends.
\end{abstract}
\centerline{PACS number: 03.65.Pm; 11.17.-w}
\bigskip
\par
In the 50-th years W.Heisenberg introduce the nonlinear term
in the Dirac  equation \cite{h1}-\cite{h3}.
The  given equation  (Heisenberg
equation(HE)) has the discrete spectrum of the spherically
symmetric solution with the finite  energy  even  in  the
classical region \cite{fin1}, \cite{fin2}.  This  gave  the
hope that in Unified
field theory  based  on  HE  all the fundamental characteristic
of the elementary particles would be derive.  The
further development of the theory straight this direction
showed  that this hope could not be realized.
\par
Notice that  the  main  peculiarity of HE is the fact that it
has the  non-local  solution:  for  example,  in
\cite{fin1},\cite{fin2} the
spherically symmetric particlable solutions are received. It
can be  suppose  that  also  the  axially  symmetric  non-local
solutions exist in HE which can be described strings or
flux tubes. Recall  that  at  present  the  strings  and  the flux
tubes  are
actively discussed in the superstring theory \cite{wit} and in
quantum chromodynamics (see for example \cite{nus}-\cite{ols}).
This paper is devoted to the string solutions in HE.
Remark  that  the  axially  symmetric string  solution has been
already found in the nonlinear Schr\"odinger equation \cite{dzh}.
\par
Let us write down the HE by the next view:
\begin{equation}
\left[i\gamma ^{\mu}D_{\mu} - m + \lambda
(\bar{\psi} \psi )\right]\psi = 0,
\label{a1}
\end{equation}
where $\bar{\psi}$ is the Dirac adjoint spinor;
$D_{\mu}=\partial /\partial x^{\mu} + iG_{\mu}$ is the covariant
derivative; $G_{\mu}$ is the Yang - Mills field;
$\lambda$ is the some coefficient;  $\hbar = c = 1$; $\mu=0,1,2,3$.
The nonlinear  term  in
Eq.(\ref{a1}) can take various forms:
$\gamma ^{\mu}(\bar{\psi}\gamma _{\mu}\psi)$,
$\gamma^{\mu}\gamma^{5}(\bar\psi\gamma _{\mu}\gamma^{5}\psi)$,
but we choose the
simpliest case represented in Eq.(\ref{a1}).  We will search
for the axially symmetric solution of the system (\ref{a1}).
For that we write down $\psi(t,\rho ,\varphi)$ as follows:
\begin{equation}
\psi (t,\rho ,\varphi) = \left(
 \begin{array}{rr}
A(\rho)& \\
B(\rho) &e^{i\varphi} \\
iC(\rho)& \\
iD(\rho) &e^{i\varphi}
 \end{array}
 \right) exp(-iEt),
\label{a2}
\end{equation}
where $\rho$, $z$ and $\varphi$ are  the  coordinates
in  the  cylindrical  coordinate system; $G_\mu = 0$, $E$ - some
constant. The substitution of (\ref{a2}) into Eq.(\ref{a1}) leads to
the next set of the equations:
\begin{eqnarray}
(E-m)A -D'-\frac{D}{\rho} + \lambda A
(A^{2} + B^{2} -C^{2} - D^{2})&=&0,\\
(E-m)B -C' + \lambda B
(A^{2} + B^{2} -C^{2} - D^{2})&=&0,\\
-(E+m)C -B'-\frac{B}{\rho} + \lambda C
(A^{2} + B^{2} -C^{2} - D^{2})&=&0,\\
-(E+m)D -A' + \lambda D
(A^{2} + B^{2} -C^{2} - D^{2})&=&0.
\label{a34}
\end{eqnarray}
\par
We consider the simpliest case $B=C=0$. We introduce the
dimensionless variable $x=\rho m$; $\beta = E/m$ and
the functions $A(\rho )~=~a(x)(m/\lambda )^{1/2}$,
$D(\rho )~=~d(x)(m/\lambda )^{1/2}$. Thus we get the next
set of equations:
\begin{eqnarray}
a' + d(1 + \beta - a^{2} + d^{2})&=&0,
\label{a41}\\
d' + \frac{d}{x} + a(1 - \beta -a^{2} + d^{2})&=&0.
\label{a42}
\end{eqnarray}
\par
As it easy to note, the given set has the following trivial
solutions:
\begin{eqnarray}
a = d&=&0,
\label{a51}\\
d=0;\qquad a&=&\pm \sqrt{1-\beta}.
\label{a52}
\end{eqnarray}
\par
The solution (\ref{a51}) is the saddle point in the phase
space $(ad)$ and the solutions (\ref{a52}) are two stables
focuses.
\par
The set (\ref{a41},\ref{a42}) coincides (up to the factor $2$
at the second summand in Eq.(\ref{a42}))
with that investigated in \cite{fin1},\cite{fin2}. On those
papers the qualitative investigation of the set of equations
for spherically symmetric case was carried out, and as a
result of it was shown that there exists the finite discrete
spectrum of the solutions making a physical sense. These solutions
asymptotically drop to zero by an exponential law and are regular
in the
whole space, so they have a finite energy. It is logical to
suppose that in the axially symmetric case this features will
take place too, that is solutions making a physically sense with
the finite energy on the unit of length will take place. Such the
solutions would be named as string ones and it would allow to get
a new approach to the HE.
\par
We shall solve the set (\ref{a41}-\ref{a42}) by numerical way.
The solution has the following form near the axis $z=0$:
\begin{eqnarray}
a&=&a_{0} + a_{2} \frac{x^{2}}{2} + \cdots,
\label{b1}\\
d&=&d_{1}x + \cdots,\\
a_{2}&=&d_{1}(a^{2}_{0} -1 - \beta), \\
d_{1}&=&\frac{1}{2} a_{0} (a^{2}_{0} -1 + \beta ).
\label{b2}
\end{eqnarray}
\par
{}From  (\ref{b1}-\ref{b2})  we see that  all the solutions  depend
upon two parameters: $a_{0}=a(x=0)$ and $\beta$.
It is easy to note that there is a continuous region of values
$a_{0}$ for which
$a(x)\stackrel{x\to \infty}{\longrightarrow}+(1-\beta )^{1/2}$,
it is the stable focus (\ref{a41}-\ref{a42})
in the solution of the set (\ref{a41}-\ref{a42}).
Moreover, there is a continuous region of values $a_{0}$ for which
$a(x)\stackrel{x\to \infty}{\longrightarrow}-(1-\beta )^{1/2}$,
it is the stable focus (\ref{a41}-\ref{a42}).
It is evident that the exceptional solution take place
on the boundary of these regions (for which
$a(x)\stackrel{x\to \infty}{\longrightarrow}0$,
$d(x)\stackrel{x\to \infty}{\longrightarrow}0$).
It is the separatrix passing through the saddle
point (\ref{a51}). Denote the boundary value of the exceptional
solution by the  following: $a(x=0)=a^{*}_{n}$, where $n$
the number of the points in which $a(x)=0$.
\par
These exceptional solutions $a_{n}(x)$ and $d_{n}(x)$ may be found
by an sequential  approximation method. Two solutions
$a_{0}(x)$, $d_{0}(x)$ and $a_{1}(x)$, $d_{1}(x)$
are displayeds on the Fig.1
($\beta=0.5$).
By the numerical method  the following boundary values are
evaluated:
$a^{*}_{0}=1.298569\ldots$; $a^{*}_{1}=1.491163\ldots$;
$a^{*}_{2}=1.576012\ldots$.
By further numerical analyze was find out that solutions
are singular by $x>2.0$.
\par
Apparently, the asymptotic behavior of the given exceptional
solutions is:
\begin{equation}
a\approx d\approx \frac{\exp \left\{-2x\sqrt{\beta
(1-\beta }\right\}}{\sqrt{x}}.
\end{equation}
It is clearly seen that such an asymptotic behavior of
the functions $a_{n}(x)$ and $d_{n}(x)$ tends to the
fact that the linear density of the energy is finite
for the given string.
\par
Now we are proceeding to the physical interpretation of the
received results. We calculate the linear energy density of the
string. According to the general Hamiltonian definition,
the linear energy density is equal:
\begin{equation}
h = \int \bar \psi \left[ -i \gamma ^j \partial _j +
m - \frac{\lambda}{2} \left(\bar \psi \psi \right)
\right] \psi \rho d\rho d\varphi,
\label{a18}
\end{equation}
where $j=1,2,3$. Then we receive the following expression for
the energy density:
\begin{equation}
h = \frac{2\pi}{\lambda} I_n,
\label{19}
\end{equation}
where
\begin{equation}
I_n = \int \limits_{0}^{\infty}\left[ \beta \left( a^2
+b^2 \right) + \frac {1}{2} \left( a^2-b^2 \right) ^2
\right] xdx.
\label{20}
\end{equation}
The numerical calculations give us the following results:
$I_0~\approx~3.5524$,
$I_1~\approx~21.4293$,
$I_2~\approx~51.7859$.
\par
Despite  the boundary effect we assume that the
quark and antiquark rotating around the common center
are placed on the ends of the string
having the linear energy density (\ref{a18}). In according
with well-known opinion this construction is a meson.
Take another frame of reference  connected with the rotating
quark, after that its
potential energy $V$ is equal:
\begin{equation}
V(r) =\frac{2\pi}{\lambda} I_n r + \frac{L^2}{2Mr^2},
\label{21}
\end{equation}
where $r$ is the distance from quark to the mass center;
$M$ is the quark mass;
$L$ is the angular momemtum of the quark. This potential
will have the minimum if only $I_n>0$. In this case the quark
will be in the potential hole and its distance $r_0$
to the rotation center will equal:
\begin{equation}
r_0 \approx \left( \frac{\lambda}{2\pi I_n M}\right) ^{1/3},
\label{22}
\end{equation}
where we take into account that the characteristic angular
momentum for meson is $L\approx \hbar = 1$. As
the suggested string model is a phenomenological model,
we cannot define the parameter $\lambda$. The
parameters $\lambda, E$ and $m$ can be define in the scope
of Yang-Mills nonperturbative quantum Yang-Mills field theory,
by analogy with that occure in superconductivity theory during
the definition of the parameters of Ginzburg-Landau equation from
the microscopic theory. It is interesting is to estimate the value
of the parameter $\lambda$ defining the quantum
self-action of the string wave function; i.e.
\begin{equation}
\lambda \approx 2\pi I_n Mr^3_0
\label{23}
\end{equation}
\par
The characteristic quark mass is $M\approx 0.5 GeV$,
characteristic meson length (string length) is
$r_0 \approx 1Fm$, hence then the
parameter $\lambda \approx 4.8 Fm^2$ with $n=0$. It is
easy to show that according to dimensional consideration
we can write down this parameter as follows
$\lambda = \hbar c l^2 = l^2$, where
$l \approx 2.2 Fm$ is a certain length scale constant.
\par
It is obvious that this consideration is an approximative
model of meson as a string stretshed  between rotating
quark-antiquark pair. As the next approximation we
have to take into account the chromodynamical field created
by the quark.
\par
Finally, we see that the HE has the non-local solutions
and in particular it has the string solution with the finite
energy on the unit of length. By analogy with the Dirac equation
describing the quantum pointlike particle we can assume that
the Heisenberg equation describe the quantum non-local objects,
for example: axially symmetric objects (strings, flux tubes),
spherically symmetric objects (membranes) and so on. The received
solution can use to the obtaining the string model of meson.

\newpage
\centerline{List of the figure captions}
\noindent
\centerline{Fig.1.}
\par
$1$ is $a_{0}(x)$ function; $2$ is $d_{0}(x)$ function;
$3$ is $a_{1}(x)$ function; $4$ is $d_{1}(x)$ function.
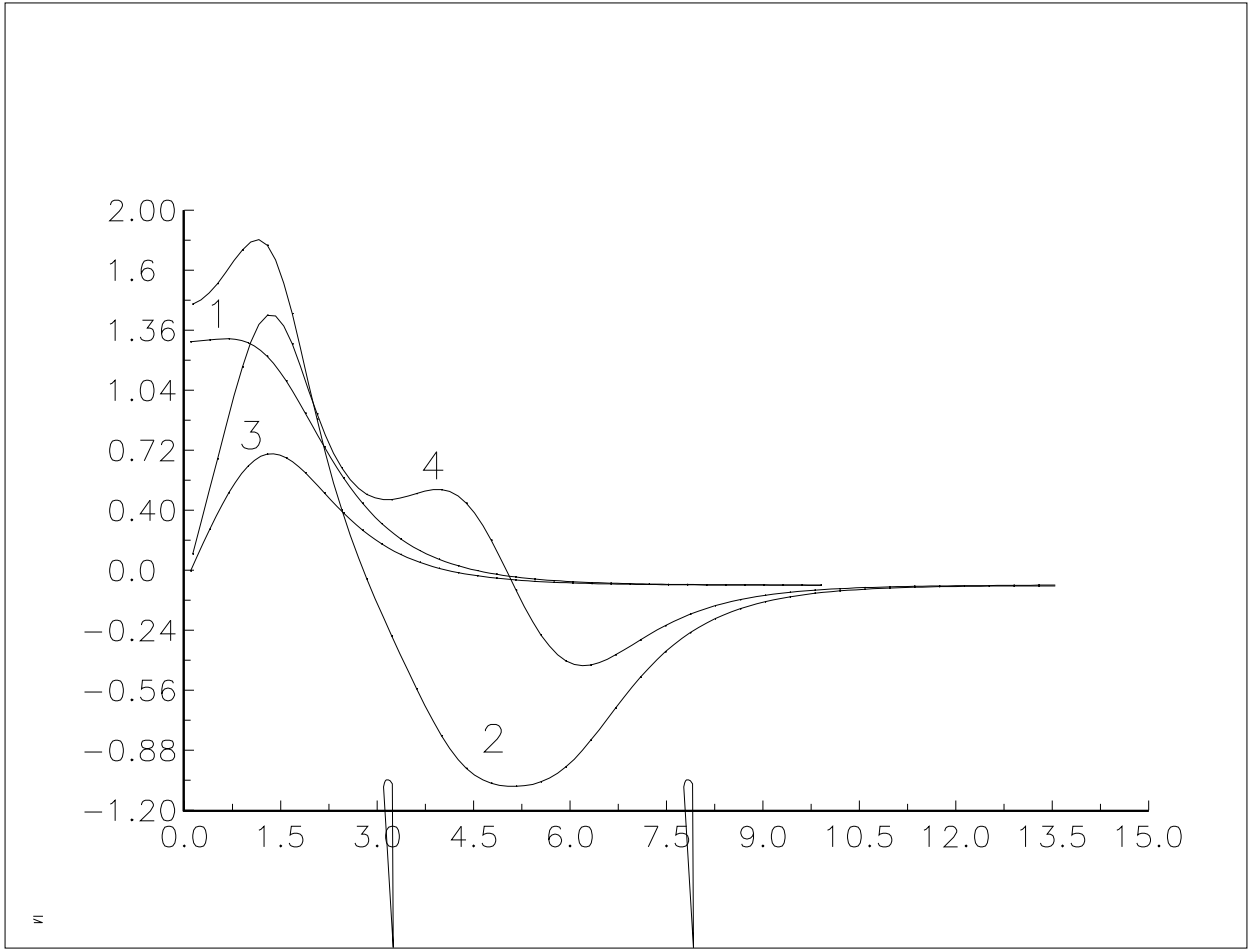
\end{document}